\documentclass[3p,times,12pt]{elsarticle}
\usepackage{url}
\usepackage[usenames, dvipsnames]{color}
\usepackage{xcolor, soul}
\sethlcolor{green}
\usepackage{array}
\usepackage{caption}
\usepackage{wrapfig}
\usepackage{amssymb}
\usepackage{booktabs}
\usepackage{graphicx}
\usepackage{tabulary}
\usepackage{subcaption}

\usepackage{color}

\newcommand{\unit}[1]{\ensuremath{\,\mathrm{#1}}}

\journal{ }

\begin{document}

\begin{frontmatter}

\title{Algorithm and Simulation of Heat Conduction Process for Design of a Thin Multilayer Technical Device}

\author[jinr-lit]{{Alexander}~Ayriyan\corref{corr}}\ead{ayriyan@jinr.ru}
\author[tuke,sinica]{{J\'{a}n}~Bu\v{s}a~Jr.}
\author[jinr-lhe]{{Eugeny}~E.~Donets}
\author[jinr-lit,ysu]{{Hovik}~Grigorian}
\author[jinr-lit,tuke]{{J\'{a}n}~Pribi\v{s}}

\address[jinr-lit]{Laboratory of Information Technologies, JINR, Dubna, Russia}
\address[tuke]{Department of Mathematics and Theoretical Informatics, TU of Ko\v{s}ice, Ko\v{s}ice, Slovakia}
\address[sinica]{{\color{black} Institute of Physics, Academia Sinica, Nankang, Taipei, Taiwan}}
\address[jinr-lhe]{Veksler and Baldin Laboratory of High Energy Physics, JINR, Dubna, Russia}
\address[ysu]{Department of Theoretical Physics, Yerevan State University, Yerevan, Armenia}

\cortext[corr]{Corresponding author}

\begin{abstract}
A model of a multilayer device with non-trivial geometrical structure and nonlinear dependencies of thermodynamic material properties at cryogenic temperatures is suggested. A considered device, called cryogenic cell, is intended for use in multicharged ion sources for pulse injection of gaseous species into ionization space of ion sources. The main requirement for the cryogenic cell operation is the permanent opening and closing for gaseous species injection in a millisecond range, while cell closing is provided by freezing of the gaseous specie at the outer surface of the cell and the cell opening -- by the corresponding pulse heating of the cell surface up to definite temperature. The thermal behaviour of the device in a millisecond time range is simulated. The algorithm for solving the non-stationary heat conduction problem with a time-dependent periodical heating source is suggested. The algorithm is based on finite difference explicit-implicit method. The OpenCL realization of the algorithm is discussed. {\color{black} The optimal particular choice of the parameters to provide the required pulse temperature regime of the designed cryogenic cell for the chosen working gas is presented. Based on {\color{black} these} results further optimization {\color{black} can be formulated}.}

\end{abstract}


\end{frontmatter}

\section{Introduction}
\label{intro}

{\color{black} In modern science the phenomena of thermal conductivity is a very common technology for {\color{black} studying} of complex objects with complex geometric and physical structure}. The main goal of this work is to suggest a model of temperature evolution for a multilayer cylindrical object, called cryogenic cell, which has application in sources of multicharged ions~\cite{donets_2010}. The function of the cryogenic cell is a pulse injection (in the millisecond range) of the working gases into the working space of the ion source.

{In the modern applied thermal engineering the dependences of physical properties of materials on temperature (even in small interval) can ensure elegant way for the diagnosis, monitoring and management of complex systems. Such interesting examples are presented in} \cite{jeong_2013} {and} \cite{franco_2014}.

Pulse injection of a gaseous species could be provided by fast mechanical gate valves, however, robust operation in millisecond range in cryogenic {\color{black} environment} is out of their possibilities.
{\color{black} The use of temperature properties of gases at cryogenic temperatures could be a real alternative to pulse mechanical gate valves in a millisecond range time operation}.
Indeed, dependency of vapor pressure of all gases on temperature is very strong in a cryogenic temperature range, i.e. between liquid helium temperature 4.2 K and liquid nitrogen temperature $78~\unit{K}$. Typical data for Krypton is given in Table~\ref{krypton} (see \cite{honig_1960}).

\begin{table}[ht]
\caption{Krypton vapour pressure as a function of temperature}
\label{krypton}
\centering
\begin{tabular}{l c c c c}
ine
Temperature, \unit{K}  &  $27.9$  &  $29.4$  &  $30.9$  &  $32.7$ \\
Krypton vapour pressure, \unit{Torr}  &  $1.3 \times 10^{-13}$  &  $1.3 \times 10^{-12}$  &  $1.3 \times 10^{-11}$  &  $1.3 \times 10^{-10}$ \\
ine
\end{tabular}
\begin{tabular}{l c c c c}
ine
Temperature, \unit{K}  &  $34.6$  &  $36.8$  &  $39.3$  &  $42.2$ \\
Krypton vapour pressure, \unit{Torr}  &  $1.3 \times 10^{-9\>\>}$  &  $1.3 \times 10^{-8\>\>}$  &  $1.3 \times 10^{-7\>\>}$  &  $1.3 \times 10^{-6\>\>}$ \\
ine
\end{tabular}
\end{table}
It is known in ion sources technology~\cite{brown_1989}, that if a working gas vapour pressure is around $10^{-6}~\unit{Torr}$, {\color{black} its} typical injection time from the injection cell into the ionization space of ion source is about 1 ms. For example, for Krypton it corresponds to the temperature {\color{black} of} $42.2~\unit{K}$. Injection cell in this case should be placed in a vicinity of working space of ion source, about 1 cm aside of ionization region of ion source. Another side, if gas vapour pressure is about $10^{-13}$ Torr it means that all gas molecules are frozen at the cell surface which has such temperature; for Krypton, for example, this temperature is $27.9~\unit{K}$. This opens possibility to use such temperature dependencies for gas injection in millisecond time range.

One needs to create such cryogenic cell which {\color{black} provides} change of its surface temperature from, say $20~\unit{K}$ up to, say, $45~\unit{K}$, and back during few milliseconds and with a {\color{black} frequency} about $50~\unit{Hz}$.

Some additional requirements for such cryogenic cell construction are inspired by some basic aspects of ion source technology and cryogenic technics:
\begin{itemize}
\item[---] there are two natural temperature terminals in cryogenics -- liquid helium temperature terminal $4.2~\unit{K}$ and liquid nitrogen temperature terminal $78~\unit{K}$, thus it is natural to use such temperature terminals as a thermostats with big capacity;
\item[---] cryogenic cell surface should be heated up by pulse electric current, passing  through conductive layer in a vicinity of cell surface; in order to prevent disturbances in a working space of ion source it has to be placed in a vicinity of a working ionization space of ion source. The~maximal electric current $I$ and  through the cell a corresponding voltage should be restricted {\color{black} to } $I \times R < 1000~\unit{V}$, where $R$ is a resistance of the conductive layer.
\end{itemize}
Design of such cryogenic cell has been elaborated and {\color{black} recently} tested in JINR \cite{donets_2012,donets_2015}. {\color{black} Which shows, that the cell exhibited expected above mentioned time and temperature parameters. For the reason of practical use in ion sources one needs to create {\color{black} a} sample of cryogenic cell which fits {\color{black} more precisely} the time and temperature requirements in millisecond time range at cryogenic temperatures}. {\color{black} The}~present work describes details of numerical strategy, {\color{black} which is} used to create suitable numerical tool for optimization of cryogenic cell construction.

\begin{figure}[h!]
\centering
\includegraphics[width=10cm]{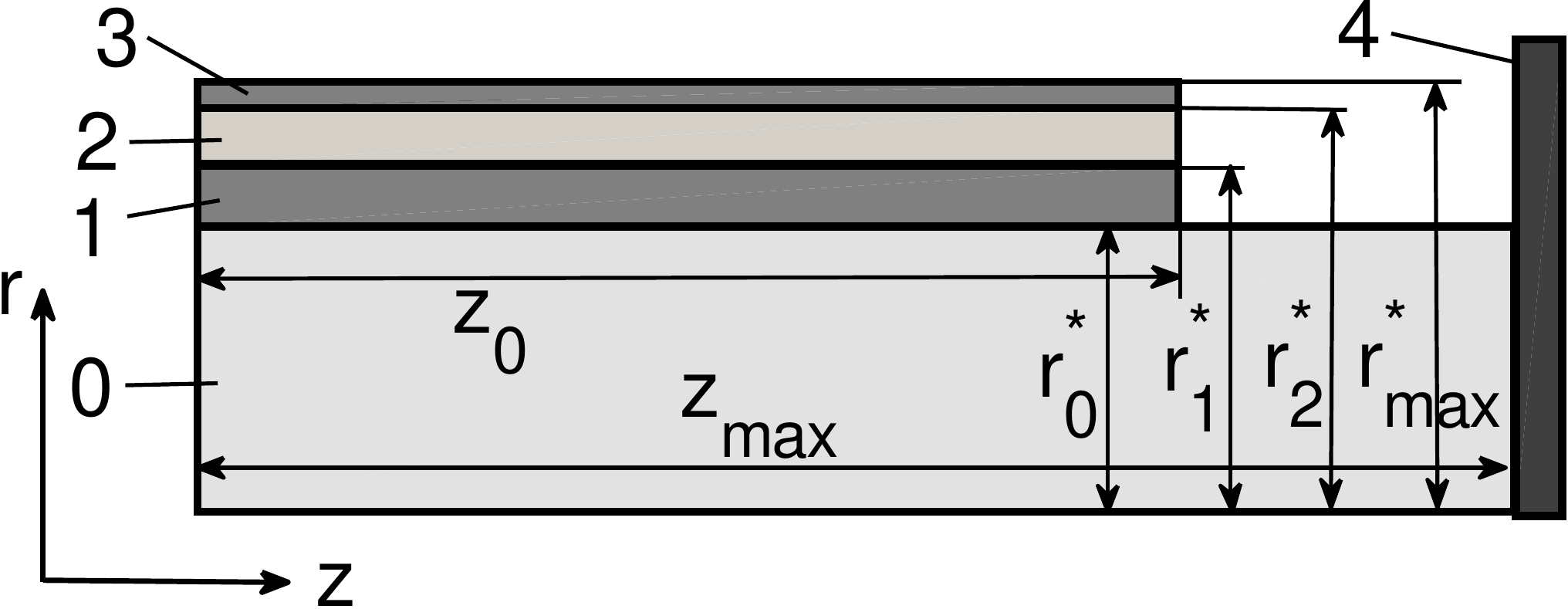}
\caption{\footnotesize  {\color{black} Schematic view of the object half-slice. The bottom line is the axis of a cylinder (axis of the symmetry), $r=0$}. The slice of the object: 0 -- cooler, 1 -- electrical insulator, 
2 -- heat source (conductive layer), 3 -- external insulator, 4 -- liquid helium temperature terminal with 
$T=4.2\unit{K}$.}
\label{fig_object}
\end{figure}

So, one needs to simulate thermal process in a cell of a chosen geometry, which is governed by the periodic passage of electric current through one of the layers of the cell. The period of the process is requested to be $t_{\mathrm{prd}}=t_{\mathrm{src}}+t_{\mathrm{clg}}$. Here $t_{\mathrm{src}}$ is {\color{black} a} period {\color{black} of} heating and $t_{\mathrm{clg}}$ is {\color{black} a} period {\color{black} of} cooling down. The period is divided in two parts: when the cell evaporates working gases from its surface ($T>T_{\mathrm{crit}}$) and when the rest of the working gas (which is not penetrated into ion source ionization space) freezes on the surface ($T<T_{\mathrm{crit}}$). The cell itself should work at the cryogenic temperature range from temperature of liquid helium ($T=4.2\unit{K}$) up to {\color{black} the} temperature of liquid nitrogen. The cell structure has a cylindrical symmetry; therefore, the heat conductivity inside it can be simulated by a model with two spatial cylindrical coordinates, $r$ and $z$, and time variables (Fig.~\ref{fig_object}). Similar but more simple model has been discussed in~\cite{ayriyan_2012_mm}~and~\cite{ayriyan_2012_lncs}.

\section{Main Equations and Boundary Conditions}
\label{model}

The thermal processes in the object can be described by the following system of
parabolic partial differential equations with temperature depended coefficients~\cite{samarski_1995}:
\begin{equation}
\label{equation}
\rho_m {c_V}_m (T)\frac{\partial T}{\partial t}=\frac{1}{r}
\frac{\partial}{\partial r} \left( r \lambda_m(T) \frac{\partial
T}{\partial r}\right ) + \frac{\partial}{\partial z} \left( \lambda_m(T) \frac{\partial
T}{\partial z}\right ) + X_m(T,t),
\end{equation}
where $r \in [0, r_{\max}(z)]$, $z \in [0, z_{\max}(r)]$ (or $\left(r,z\right)\in\Omega$) and $t \geq 0$.
The object consists of different materials in construction with different densities and thermal coefficients; thus, the index $m$ is introduced for each material ($m = 0$ -- cooler (copper), $m = 1$ -- electrical insulator, $m = 2$ -- heat source (graphite), $m = 3$ -- external insulator). In the frame of this work, physical and engineering needs of the object geometry are not discussed.
The source function in Eq.~(\ref{equation}) is $X_m(T)\equiv 0$ for the layers $m=0$,$1$, and $3$ (there is no source) and {\color{black} it} has a periodical time dependence:
\begin{equation}
\label{source}
{\color{black} X_2(T,t) = \chi(T) \frac{I^{2}(t)}{S_2^2},}
\end{equation}
{\color{black} where $S_2$ is the cross-sectional area in units of $\!\unit{cm^{2}}$ and $\chi(T)$ is temperature depended resistivity of the conducted layer $m=2$.
The current can be expressed in the form:}
\begin{equation}
\label{current}
\displaystyle I(t)=I_{0}\nu(t)p(t),
\end{equation}
{\color{black} where $I_{0}$ is the amplitude of current and $\nu(t)$ represents the time structure:}
\begin{equation}
\label{period}
\nu(t) = \left\{
\begin{array}{l l}
    1, & \! nt_\mathrm{prd} \leq t < nt_\mathrm{prd} + t_\mathrm{src},\\
    0, & \! nt_\mathrm{prd} + t_\mathrm{src} \leq t < (n+1)t_\mathrm{prd},
\end{array} \right.
\end{equation}
here $n\in\mathbb{N}_{0}$ {\color{black} -- is index of a period of the electric current.}
Function $\nu(t)$ has a uniform rectangular waveform ({definition of this function one can find in} ~\cite{symons_2013}), and $p(t)$ is a model of the transient response function for the turn-on process ({it is introduced in analogous with similar function in electrical engineering, see for example}~ \cite{gajic_2002}). In a simple case, it can be the Heaviside step function. For this work the function $p(t)$ is discussed in Section~\ref{tproc}. In the formulas, $I(t)$ stands for electric current in the graphite slice along $z$ direction. In a common case, the thermal coefficients are nonlinear functions of the temperature and the spatial coordinates with discontinuities of the first kind (for $m+1$ materials, there are $m$ points of discontinuities).

The initial condition is given by
\begin{equation}
\label{initcond} 
T(r,z,t=0) = T_0,
\end{equation}
\noindent where $T_0\equiv 4.2\unit{K}$ (liquid helium temperature) and the boundary conditions are taken as
\begin{equation}
\label{boundcond}
\left\{
\begin{array}{l}
\displaystyle \frac{\partial T}{\partial \mathbf{n}} =0 \qquad \forall \, (r,z) \in \delta\Omega \setminus \lbrace (r,z): z=z_\mathrm{max} \rbrace,\\[3mm]
\displaystyle T=T_{0} \qquad \forall \, (r,z) \in \lbrace (r,z): z=z_\mathrm{max} \rbrace,
\end{array}
\right.
\end{equation}
where $\delta\Omega$ is the boundary of the $\Omega$, and $\mathbf{n}$ is the normal vector of $\delta\Omega$. The temperature at the right side is always equal to $T_{0}$ because of contact with liquid helium.

The parameter $\rho$ and the functions ${c_V}$, $\lambda$, and $X_{i}=X(T_i)$ have discontinuities of the first kind at the following surfaces with radii: $r_0^*$, $r_1^*$, and $r_2^*$ in the interval $[0,r_{\max}]$. Conjugation conditions between materials are considered to be ideal:
\begin{equation}
\label{sch}
\left\{
\begin{array}{l}
\displaystyle \left.T\right|_{r=r_m^* -0} = \left.T\right|_{r=r_m^* +0},\\[2mm]
\displaystyle -\lambda_{m}(T)\left.\frac{\partial T}{\partial r}\right|  _{r=r_m^* -0} = -\lambda_{m+1}(T)\left.\frac{\partial T}{\partial r}\right|  _{r=r_m^* +0},
\end{array}
\right.
\end{equation}
where $r_{m}^*$ are points of the border between the materials $m$ and $m+1$ (discontinuity points), here $m=0,1,2$.

\section{Numerical Algorithm}
\label{algo}

The initial-boundary value problem Eqs.~(\ref{equation})--(\ref{boundcond}) has been approximated by the following mixed explicit-implicit finite difference scheme ({see} \cite{samarski_2001, yanenko_1967}):
\begin{equation}
\label{scheme_eq}
\rho_{i,j}\,{c_V}_{i,j}\frac{\widehat{T}_{i,j}-T_{i,j}}{\tau}=
{\textrm{\Large{$\Lambda_i$}}}\left[\right.\widehat{T}_{i,j}\left.\right]+
{\textrm{\Large{$\Lambda_j$}}}\left[\right.T_{i,j}\left.\right]+
X_{i,j},
\end{equation}
where $\widehat{T}_{i,j}$ is the temperature on the next time step, $T_{i,j}$ is the temperature on the current time step, $\tau$ is the time-step.

Numerical solution of Eq.~(\ref{scheme_eq}) can be obtained using a special non-uniform grid:
\vspace{-2mm}
\begin{eqnarray}
\label{uniform_grid}
\overline{\omega}& = \lbrace (t,x,z)\left|\right. & 0 \leq t < \infty,\quad t_i = k\cdot h_t,\quad k \in \mathbb{N}_{0};\nonumber\\
                 & & 0 \leq r \leq r_{\mathrm{max}},\quad r_{i+1} = r_i+h_{i+1},\quad i = 0,\ldots, N_j-1;\\
                 & & 0 \leq z \leq z_{\mathrm{max}},\quad z_{j+1} = z_j+\eta_{j+1},\quad j = 0, \ldots, M_i-1\nonumber
\rbrace.
\end{eqnarray}

The spatial finite difference operator is:
\begin{equation}
\label{Lambda_i}
{\textrm{\Large{$\Lambda_i$}}}\left[\right.\widehat{T}_{i,j}\left.\right]=
\frac{1}{r_i}\frac{1}{\hbar_{i}}\left[r_{i+\frac{1}{2}} \lambda_{i+\frac{1}{2},j} 
\frac{\widehat{T}_{i+1,j}-\widehat{T}_{i,j}}{h_{i+1}}-r_{i-\frac{1}{2}} \lambda_{i-\frac{1}{2},j}
\frac{\widehat{T}_{i,j}-\widehat{T}_{i-1,j}}{h_{i}}\right],
\end{equation}
\def\bareta{\eta\kern-0.8ex\lower0.2ex\hbox{\vrule height0.4pt width0.8ex}\kern0.1ex}
\begin{equation}
\label{Lambda_j}
{\textrm{\Large{$\Lambda_j$}}}\left[\right.T_{i,j}\left.\right]=
\frac{1}{\bareta_j}\left[\lambda_{i,j+\frac{1}{2}} 
\frac{T_{i,j+1}-T_{i,j}}{\eta_{j+1}}-\lambda_{i,j-\frac{1}{2}}
\frac{T_{i,j}-T_{i,j-1}}{\eta_{j}}\right],
\end{equation}
where $i = 1,\ldots,N_j-1$, $j = 1,\ldots, M_i-1$, $h_{i} = r_{i}-r_{i-1}$, $\eta_{j} = z_{j}-z_{j-1}$, $\hbar_{i}=\left(h_{i+1}+h_{i}\right)/2$, $\bareta_j=\left(\eta_{j+1}+\eta_{j}\right)/2$, $\displaystyle T_{i,j}=T(r_i,z_j,t_k)$, $\displaystyle \widehat{T}_{i,j}=T(r_i,z_j,t_{k+1})$, $\displaystyle \lambda_{i,j}=\lambda_m(T_{i,j})$, $\displaystyle {c_V}_{i,j}={c_V}_m(T_{i,j})$, \linebreak $\displaystyle X_{i,j}=X_m(T_{i,j})$, $\displaystyle r_{i\pm\frac{1}{2}}=(r_i + r_{i\pm1})/2$, $\displaystyle \lambda_{i\pm \frac{1}{2},j}=\lambda_m(T_{i,j}+T_{i\pm1,j})/2$, $\displaystyle \lambda_{i,j\pm \frac{1}{2}}=\lambda_m(T_{i,j}+T_{i,j\pm1})/2$.

The indices $i$ and $j$ are global for the whole computational domain. The index $m$, which marks the material, is chosen correspondingly to the domain region, from which the pair ($i$, $j$) is taken.

The coefficients for the forward sweep of the Thomas algorithm {described in} \cite{thomas_1949, press_2007} are defined as:
\begin{equation}
\label{front_sweep}
\left\{
\begin{array}{l}
\displaystyle \alpha_{i} = \frac{-C_{i}}{B_{i}+A_{i}\alpha_{i-1}},\\[4mm]
\displaystyle \beta_{i} = \frac{F_{i}-A_{i}\beta_{i-1}}{B_{i}+A_{i}\alpha_{i-1}}.\\[1mm]
\end{array}
\right.
\end{equation}

The coefficients $A_i$, $B_i$, $C_i$, and $F_i$ are formulated from the difference equation~(\ref{scheme_eq}):
\begin{equation}
\label{ABC}
\left\{
\begin{array}{l}
\displaystyle A_{i} = \frac{-r_{i-\frac{1}{2}}\lambda_{i-\frac{1}{2},j}}{r_i\,\hbar_{i}\,h_{i}},\\[4mm]
\displaystyle B_{i} = \frac{1}{r_i\hbar_{i}}\left[ \frac{r_{i-\frac{1}{2}} \lambda_{i-\frac{1}{2},j}}{h_{i}} + \frac{r_{i+\frac{1}{2}} \lambda_{i+\frac{1}{2},j}}{h_{i+1}} \right] + \frac{\rho_{i,j}\,{c_V}_{i,j}}{\tau},\\[3mm]
\displaystyle C_{i} = \frac{-r_{i+\frac{1}{2}}\lambda_{i+\frac{1}{2},j}}{r_i\,\hbar_{i}\,h_{i+1}},\\[4mm]
\displaystyle F_{i} = \frac{\rho\,{c_V}_{i,j}}{\tau}T_{i,j} + \Lambda_j\left[\right.T_{i,j}\left.\right] + X_{i,j}.\\[1mm]
\end{array}
\right.
\end{equation} 
Finally, the following formula has been used to calculate unknown values of $\widehat{T}_{i,j}$:
\begin{equation}
\label{back_sweep}
\widehat{T}_{i,j} = \alpha_i\widehat{T}_{i+1,j} + \beta_i.
\end{equation}

For the forward sweep Eqs.~(\ref{front_sweep})--(\ref{ABC}) and back sweep Eq.~(\ref{back_sweep}), one needs to initialize values $\alpha_0$, $\beta_0$, and $\widehat{T}_{N_j,j}$ respectively. These values have to be initialized to satisfy the boundary conditions.

{It is known} ({see} \cite{samarski_1995,samarski_2001}) that in the case of a non-uniform grid or discontinuities of the first kind of the thermal coefficients, Scheme (\ref{scheme_eq}) has the first order difference approximation via spatial coordinates\footnote{Actually, the order of difference approximation depends on the choice of a norm, {as it is shown in given references}.}. Therefore, for approximations of the boundary and conjugation conditions, it is enough to use the difference approximation of the first order.

The boundary conditions have been approximated by the following set of formulas:
\begin{equation}
\label{first_oder_bc}
\left\{
\begin{array}{l}
\displaystyle \widehat{T}_{0,j} = \widehat{T}_{1,j},\\[3mm]
\displaystyle \widehat{T}_{N_j,j} = \widehat{T}_{N_j-1,j},
\end{array}
\right.
\end{equation}
for $r=0$ and $r=r_{\max}$, respectively. Difference approximations of the other boundary conditions can be constructed in a similar way.

The initial values of $\alpha$ and $\beta$ satisfying the boundary conditions (\ref{first_oder_bc}) are defined by:
\begin{equation}
\label{alphabeta_ini}
\displaystyle \alpha_0 = 1,\ \ \beta_0 = 0.\\[1mm]
\end{equation}

To initialize the recursive formula (\ref{back_sweep}), taking into account the boundary conditions from Eq.~(\ref{first_oder_bc}), the $T_{Nr,j}$ is calculated as:
\begin{equation}
\label{T_ini_for_back_sweep}
\widehat{T}_{N_j,j} = \frac{\beta_{N_j-1}}{1-\alpha_{N_j-1}}.
\end{equation}

The relations on the border between layers are given in the form
\begin{equation}
\label{cc_first_1}
\left\{
\begin{array}{l}
\displaystyle \widehat{T}_{i^*-0,j} = \widehat{T}_{i^*+0,j},\\[4mm]
\displaystyle -\lambda_{m}(T_{i^*-0,j})\frac{T_{i^*,j}-T_{i^*-1,j}}{h_{i^*}} = -\lambda_{m+1}(T_{i^*+0,j})\frac{T_{i^*+1,j}-T_{i^*+0,j}}{h_{i^*+1}}.
\end{array}
\right.
\end{equation}
Here we use the following notation: $\widehat{T}_{i^*,j}=\widehat{T}_{i^*-0,j}=\widehat{T}_{i^*+0,j}$ and $\lambda_m^* = \lambda_{m}(T_{i^*,j})$. Instead of the recursive formula Eq.~(\ref{back_sweep}), after a simple transformation, a similar function can be expressed at the discontinuity point as:
\begin{equation}
\label{cc_first_2}
\widehat{T}_{i^*,j} = \alpha_{i^*}\widehat{T}_{i^*+1,j}+\beta_{i^*}.
\end{equation}
From Eq.~(\ref{cc_first_1}), the coefficients for the forward sweep of the Thomas algorithm $\alpha_{i^*}$ and $\beta_{i^*}$ can be represented as follows:
\begin{equation}
\label{cc_first_3}
\left\{
\begin{array}{l}
\displaystyle \alpha_{i^*} = \frac{\lambda_{m+1}^*h_{i^*}}{\lambda_{m+1}^*h_{i^*}+\lambda_{m}^*h_{i^*+1}(1-\alpha_{i^*-1})},\\[4mm]
\displaystyle \beta_{i^*} = \frac{\lambda_{m}^*h_{i^*+1}\beta_{i^*-1}}{\lambda_{m+1}^*h_{i^*}+\lambda_{m}^*h_{i^*+1}(1-\alpha_{i^*-1})}.
\end{array}
\right.
\end{equation}

The difference scheme (\ref{scheme_eq})--(\ref{cc_first_3}) has unconditional stability related to spatial step $h_i$ and conditional stability related to $\eta_j$ ({see} \cite{yanenko_1967}):
\begin{equation}
\displaystyle \tau \le \frac{\min\left| \eta^2_j \right|}{2} \min\left| \frac{\rho(T,r,z) {c_V}(T,r,z)}{\lambda(T,r,z)} \right|.
\label{stability_condition}
\end{equation}

Generally, the conditional stability could be a strong limitation for the practical usage of a difference scheme. However, our scheme is practical in the cases, where spatial step is sufficiently large in one direction; moreover, where the step in other direction is too small; and that is the case of our model.

Of course, one could discuss the Alternating Direction Implicit (ADI) method ({descibed in} \cite{samarski_1995, press_2007, peaceman_1955}) which is absolutely stable with relation to the choice of spatial step. The motivation for our choice of method derives from the relative simplicity of the technical realization when compared to ADI and the natural affinity for parallel computing (parallelization in direction when it has conditional stability).

\subsection{Model of Transient Process for Source Term}
\label{tproc}

Generally, for implicit difference schema with coefficients depending on the sought-after function (temperature), one needs to make iterations for evaluation of temperature at each time step (for example, using the Newton method) \cite{samarski_1995}.

\begin{figure}[ht!]
  \centering
  \begin{subfigure}[b]{0.5\textwidth}
    \includegraphics[width=\textwidth]{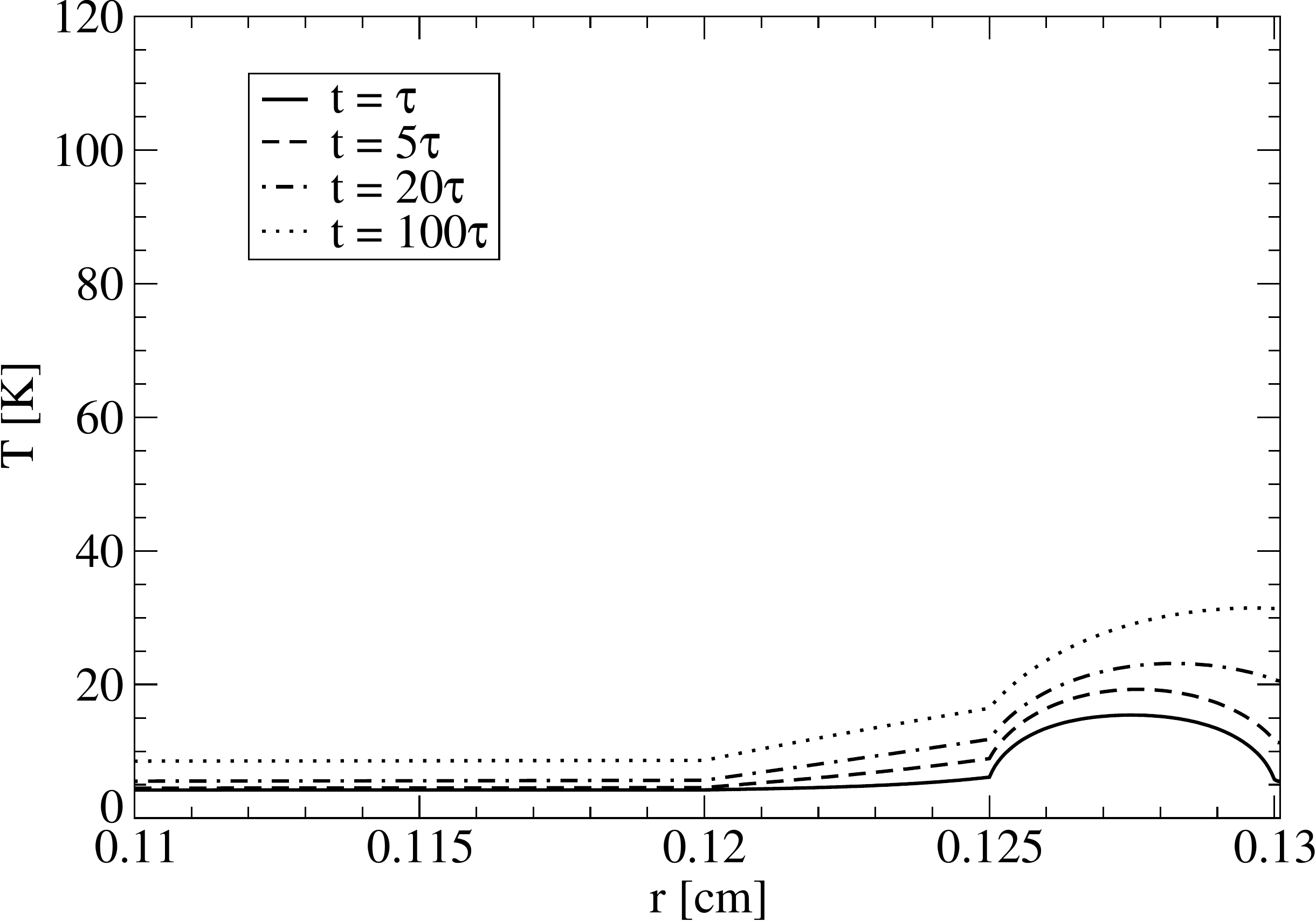}
    \caption{\footnotesize with the {\color{black}transient} model}
    \label{tr_model:with}
  \end{subfigure}%
  \begin{subfigure}[b]{0.5\textwidth}
    \includegraphics[width=\textwidth]{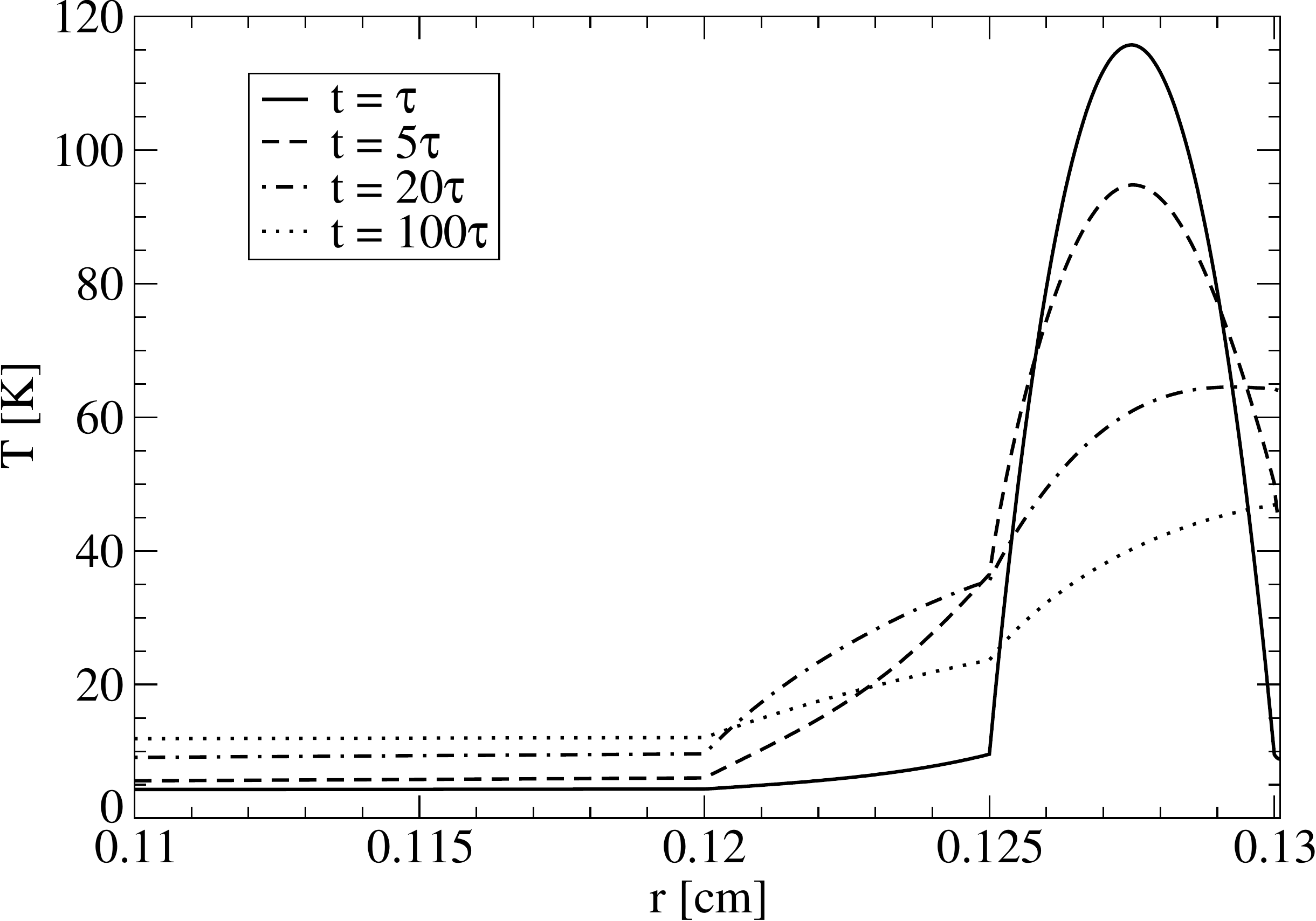}
    \caption{\footnotesize without the {\color{black}transient} model}
    \label{tr_model:without}
  \end{subfigure}
  \caption{\footnotesize {\color{black}Comparison calculations with transient model and without it for $\tau$, $5\tau$, $20\tau$, and $100\tau$, where $\tau$ is the time step of the difference scheame.}}
  \label{tr_model:both}
\end{figure}

Without using iterations, implementation of the Schema (\ref{scheme_eq})--(\ref{cc_first_3}) at the moment of noncontinuous switch on of the source (like a theta function) can lead to a jump of the temperature depending on the amplitude of the source (see Fig.~\ref{tr_model:without}). Of course, the choice of sufficiently small time-step will avoid that problem without using iterations. However, we introduce another more physically motivated method to solve this problem, namely the transient process, to avoid {\color{black} sharp behavior of temperature at the switching on of the source}.

\begin{wrapfigure}{r}{0.5\textwidth}
 \vspace{-5mm}
 \begin{center}
        \includegraphics[width=0.45\textwidth]{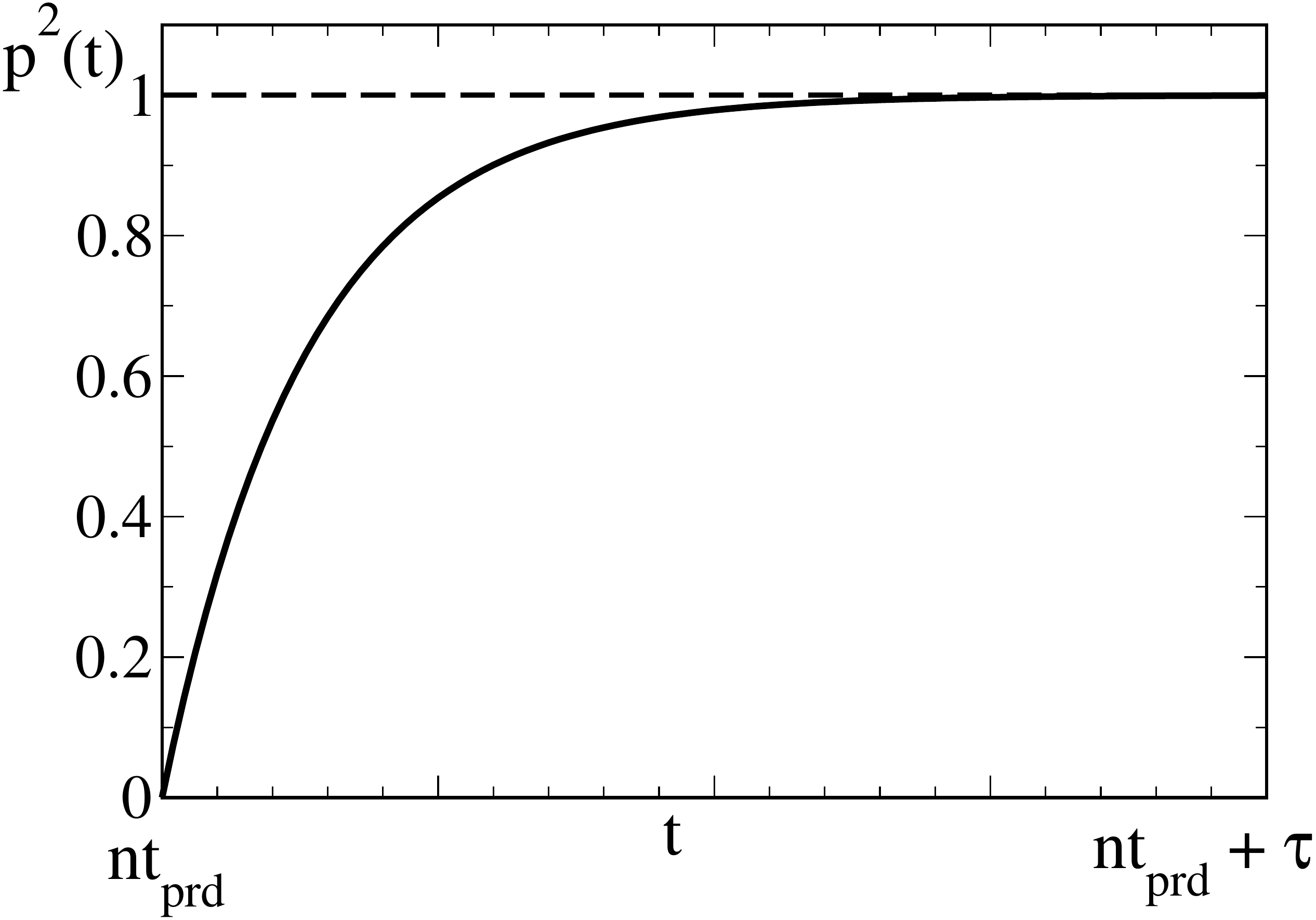}
 \vspace{-5mm}
 \end{center}
\caption{\footnotesize Transient model function $p^2(t)$.}
\label{fig:transient}
\end{wrapfigure}

Formally, it can be done with a choice of the function (Fig.~\ref{fig:transient})
\begin{equation}
\label{transient_function}
p^2(t) = \left\{
\begin{array}{l l}
    1-\mathrm{exp}\left(\displaystyle \frac{t-nt_\mathrm{prd}}{7\tau^*}\right), & \!\! nt_\mathrm{prd} \leq t < nt_\mathrm{prd} + \tau,\\
    1, & \!\! \textrm{otherwise,}
\end{array} \right.
\end{equation}
instead of $p^{2}(t)\equiv1$. Here $\tau^*=\tau/100$ and the number $7$ is introduced to have saturation occurring approximately in the middle of the time interval $\tau$ ($7$ is a magic number in relation to $100$). {\color{black} Note that we choose this function to be suitable for numerical calculation.}

The suggested of transient process is simulated in the following way. When the source is switched on (see Eq.~(\ref{period})), it starts at each moment $t = n t_{\mathrm{prd}}$. We performe the calculations with time-step $\tau^*$ up to $t= n t_{\mathrm{prd}}+\tau$. Further, calculations continue with time-step $\tau$ and $p^2(t)\equiv1$ up to $t= (n+1) t_{\mathrm{prd}}$. Then the transient process repeats.

In Fig.~\ref{tr_model:both}, comparison between two calculations made with (left figure) and without (right figure) the transient model is shown. Even the behavior of the solutions is completely different. In the incorrect solution (without transient model, Fig.~\ref{tr_model:without}), the temperature has a big jump and then the system starts cooling down instead of monotonic heating, which is the correct solution (with the transient model, Fig.~\ref{tr_model:with}). The implementation of the transient model method shows that this model is very economical.

\section{OpenCL Realization of the Algorithm}
\label{opencl}

\begin{figure}[ht!]
\begin{center}
\includegraphics[width=7cm]{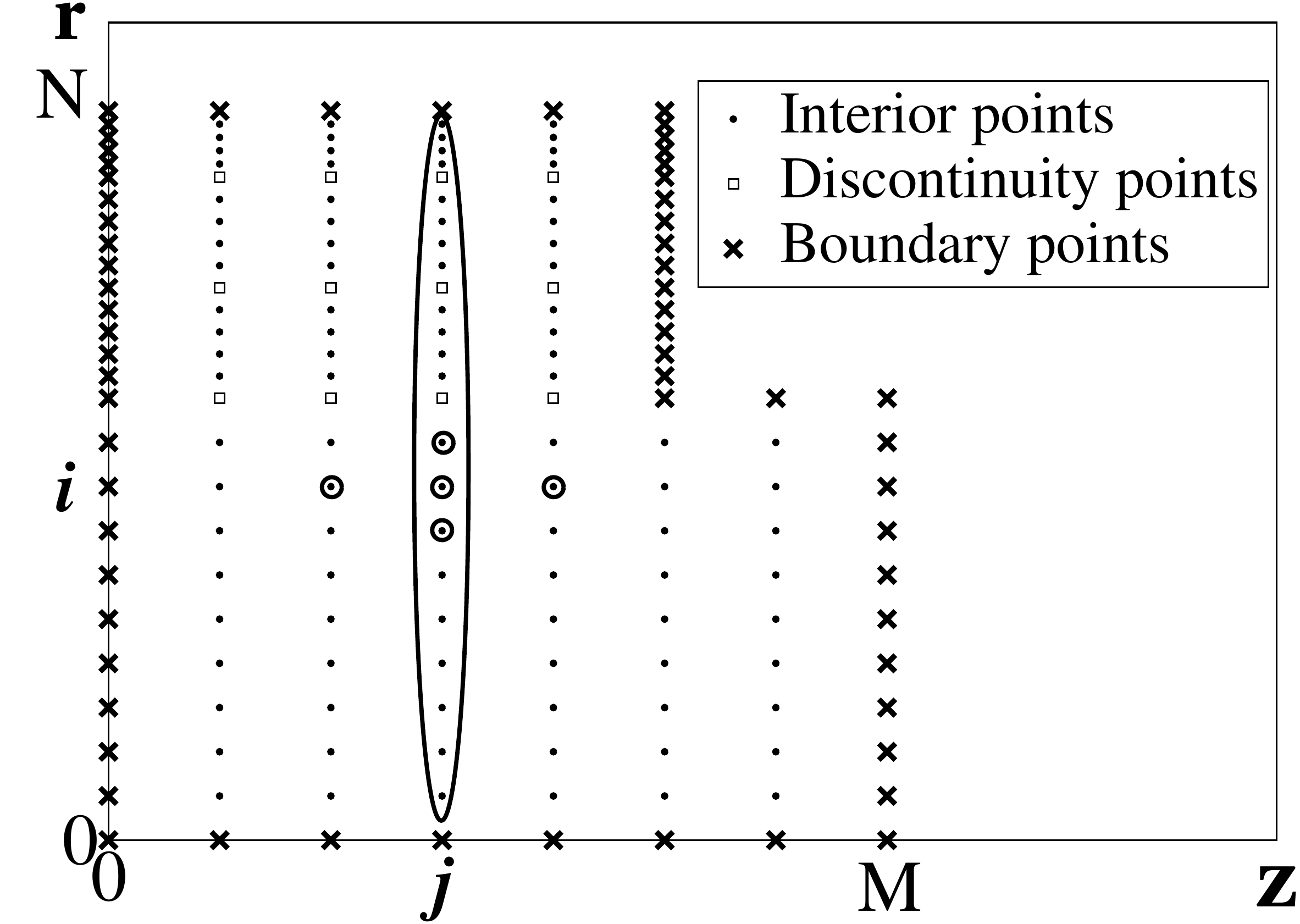}
\vspace{-5mm}
\end{center}
\caption{\footnotesize Schematic representation of discretization of the function domain.}
\label{Scheme_of_the_Algorithm}
\end{figure}

The OpenCL realization of the numerical algorithm, described in the previous section, is based on the following idea. In each time step, the cycle for $j$-index from $1$ to $M-1$ is parallelized. Each called thread simultaneously calculates the sought-after function by the Thomas algorithm, see Fig.~\ref{Scheme_of_the_Algorithm}. In the figure, we show the discretization of the function domain. Particularly, we group a~set of~points corresponding to one $j$\textsuperscript{th} thread. We also show the points involved in the calculation of the given $\left(i,j\right)$ point (circled point and crossed points on the Fig.~\ref{Scheme_of_the_Algorithm}).

\section{Results and Discussion}
\label{result}

In this work we discuss the results of the numerical simulations only for one particular choice of the configuration of the object (Fig.~\ref{fig_object}). Its geometrical characteristics have been taken as follows: $r_0 = 0.12\unit{cm}$, $r_1 = 0.125\unit{cm}$, $r_2 = 0.13\unit{cm}$, $r_{\mathrm{max}} = 0.1301\unit{cm}$, $z_0 = 4\unit{cm}$, $z_{\mathrm{max}} = 5\unit{cm}$.

\begin{figure}[ht!]
  \centering
  \begin{subfigure}[b]{\textwidth}
    \centering
	\includegraphics[width=13cm,height=5cm]{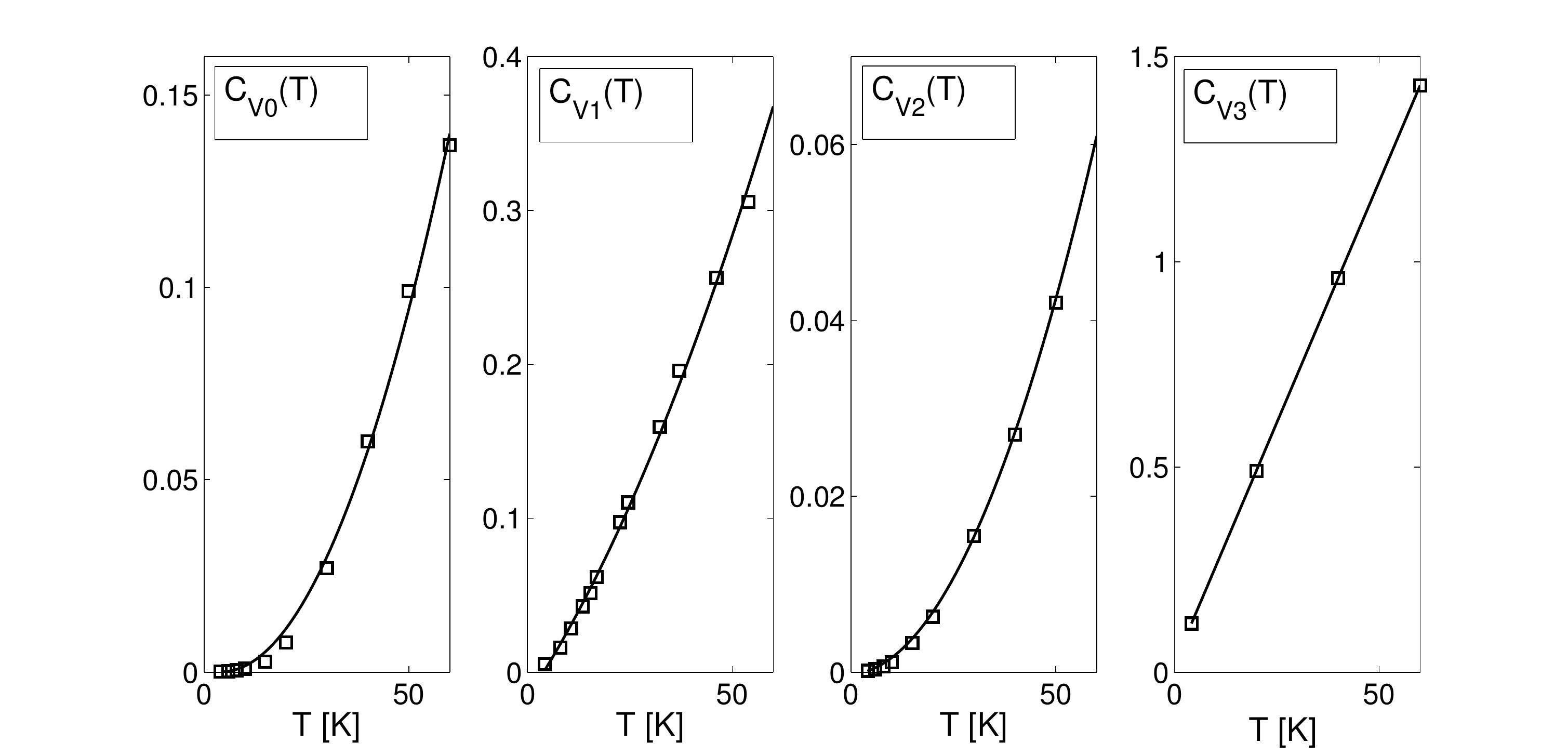}
	\caption{\footnotesize The heat capacities for different materials, $c_V(T)~\left[\unit{\displaystyle J/(g\cdot K)}\right]$. {\color{black} Squares~are data points and solid~lines~are fit~functions~Eq.~(\ref{poly_cap}).}}
	\label{fig_Cv}
  \end{subfigure}
\\
  \begin{subfigure}[b]{\textwidth}
    \centering
	\includegraphics[width=10cm]{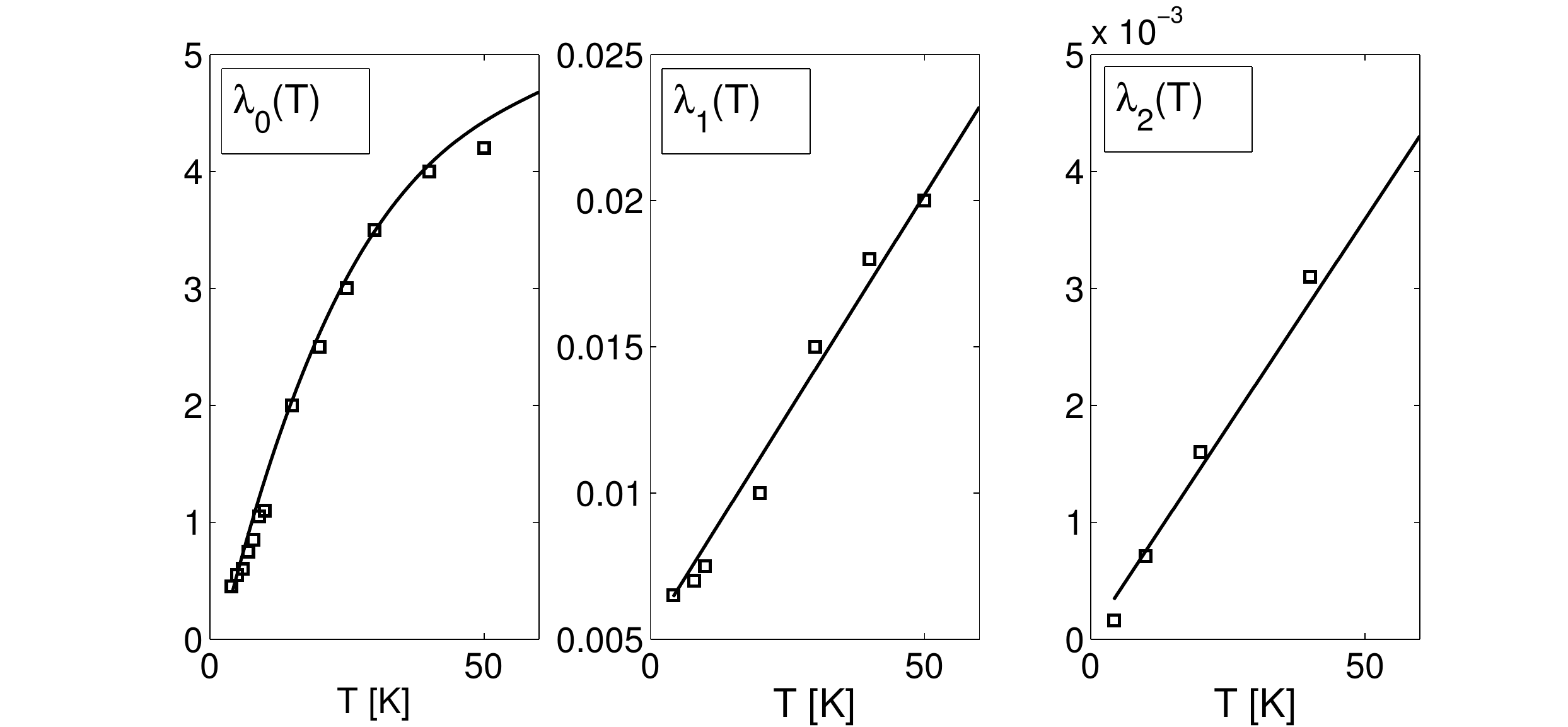}
	\caption{\footnotesize The thermal conductivities for different materials, $\lambda(T)~\left[\unit{\displaystyle W/(cm\cdot K)}\right]$. {\color{black} Squares~are data points and solid~lines~are fit~functions~Eq.~(\ref{poly_cond}).}}
	\label{fig_lambda}
	\end{subfigure}
  \caption{\footnotesize Temperature dependencies of thermal coefficients.}
  \label{tr_model}
\end{figure}

The temperature dependencies of thermal coefficients, ${c_V}_m$ specific heat capacity and $\lambda_m$ thermal conductivity, for each materials are given as it is shown in Fig.~\ref{tr_model}.
For the chosen materials the corresponding data points have been taken from~\cite{nist}. These dependencies have been fitted by the least-squares method using polynomial functions{\color{black}:\vspace{-4mm}
\begin{eqnarray}
\label{poly_cap}  {c_V}_m(T)   &=& \sum_{k=0}^{3} a_{k}^{m}\,T^{k} \\
\label{poly_cond}  \lambda_m(T) &=& \sum_{k=0}^{4} b_{k}^{m}\,T^{k}
\end{eqnarray}
The fit coefficients $a_{k}^{m}$ and $b_{k}^{m}$ are presented in the Table~\ref{a_coefs} and Table~\ref{b_coefs} correspondingly for each material $m$. The fit functions of thermal coefficients can be used for the temperature range from $4.2~\unit{K}$ up to $60~\unit{K}$.}
\begin{table}[ht]
{\color{black}
\begin{center}
\caption{{\color{black} The fit coefficients of heat capacity ${c_V}_m(T)$~Eq.~(\ref{poly_cap}).}}
\label{a_coefs}
\begin{tabular}{ l l l l l}
ine
$m\;$  &   $\;\;\, a_3$  &   $\;\;\, a_2$  &   $\;\;\, a_1$  &   $\;\;\, a_0$      \\
\toprule
$0\;$ & $\enspace\, 0.0$        & $\enspace\, 4.541 \times 10^{-4}$ & $-3.821 \times 10^{-3}$           & $\enspace\, 10^{-2}$ \\
$1\;$ & $-6.514 \times 10^{-7}$ & $\enspace\, 1.057 \times 10^{-4}$ & $\enspace\, 2.366 \times 10^{-2}$ & $-6.619 \times 10^{-3}$ \\
$2\;$ & $-8.199 \times 10^{-7}$ & $\enspace\, 2.328 \times 10^{-5}$ & $-1.245 \times 10^{-4}$           & $\enspace\, 2.169 \times 10^{-4}$ \\
$3\;$ & $\enspace\, 0.0$        & $\enspace\, 0.0$                  & $\enspace\, 2.35 \times 10^{-2}$  & $\enspace\, 2.0 \times 10^{-2}$ \\
ine
\end{tabular}
\end{center}
}\vspace{-7mm}
\end{table}
\begin{table}[ht]
{\color{black}
\begin{center}
\caption{{\color{black} The fit coefficients of thermal conductivity $\lambda_m(T)$~Eq.~(\ref{poly_cond}).}}
\label{b_coefs}
\begin{tabular}{ l l l l l l}
ine
$m\;$  &   $\;\;\, b_0$  &   $\;\;\, b_1$  &   $\;\;\, b_2$  &   $\;\;\, b_3$  &   $\;\;\, b_4$   \\
\toprule
$0\;$ & $\enspace\, 7.029\times 10^{-7}$ & $-9.317\times 10^{-5}$          & $\enspace\, 2.306\times 10^{-3}$ & $\enspace\, 1.166\times 10^{-1}$ & $-1.139\times 10^{-1}$ \\
$1\;$ & $\enspace\, 0.0$                 & $-2.523\times 10^{-7}$          & $\enspace\, 2.038\times 10^{-5}$ & $-1.248\times 10^{-4}$           & $\enspace\, 6.77\times 10^{-3}$ \\
$2\;$ & $\enspace\, 0.0$                 & $\enspace\, 3.03\times 10^{-9}$ & $-6.369\times 10^{-7}$           & $\enspace\, 1.05\times 10^{-4}$  & $-2.729\times 10^{-4}$ \\
$3\;$ & $\enspace\, 0.0$                 & $\enspace\, 0.0$                & $\enspace\, 0.0$                 & $\enspace\, 0.0$                 & $\enspace\, 10^{-3}$ \\
ine
\end{tabular}
\end{center}
}\vspace{-5mm}
\end{table}

{\color{black} The least-squares fitting of the conducted layer resistivity gives:
\begin{equation}
\chi(T)=\frac{1.8}{\sqrt{T}}~\unit{Ohm \cdot cm}.
\end{equation}} For~the external insulator (fourth material: $m=4$), we took the thermal conductivity to be constant $\lambda_3 = 10^{-3}~{\color{black}\unit{\displaystyle W/(cm\cdot K)}}$. The densities of the chosen materials are $\rho_0=8.92$, $\rho_{1}=\rho_{2}=2$ and $\rho_3=2.5$ in units $\unit{g/cm^3}$.

It is easy to see that thermal coefficients vary up to a few orders of magnitude as temperature varies in cryogenic diapason. It makes the choice of the time-step more sensitive to the values of temperature (sought-after function) (see condition~(\ref{stability_condition})), for this concrete configuration, the suitable time-step is $\tau = 10^{-5}\unit{s}$. The period of source switching is $t_{\mathrm{prd}}=25\unit{ms}$, where the heating time is $t_{\mathrm{src}}=1\unit{ms}$ -- Eq.~(\ref{period}), and electric current amplitude is {\color{black} $I_{0}=363~\unit{mA}$} -- Eq.~(\ref{current}). The critical value of temperature is taken as $T_{\mathrm{crit}}=42.2\unit{K}$ (temperature of evaporation of working gases).
The initial temperature has been taken to be equal $T_0=4.2\unit{K}$.

Note that due to the structural features of the object, especially due to the existence of tiny layers covering the core cylinder, the choice of spatial step in the radial direction has to be smaller in comparison to the size of the layers (at least in order of magnitude) to guarantee the stability of the solution.
Therefore, our choice of the difference scheme (see Section~\ref{algo}), which is suitable for the technical realization, is justified.

\subsection{Results of Numerical Simulations}
\label{modeling}

\begin{figure}[ht!]
  \centering
  \begin{subfigure}[b]{0.45\textwidth}
    \includegraphics[width=\textwidth]{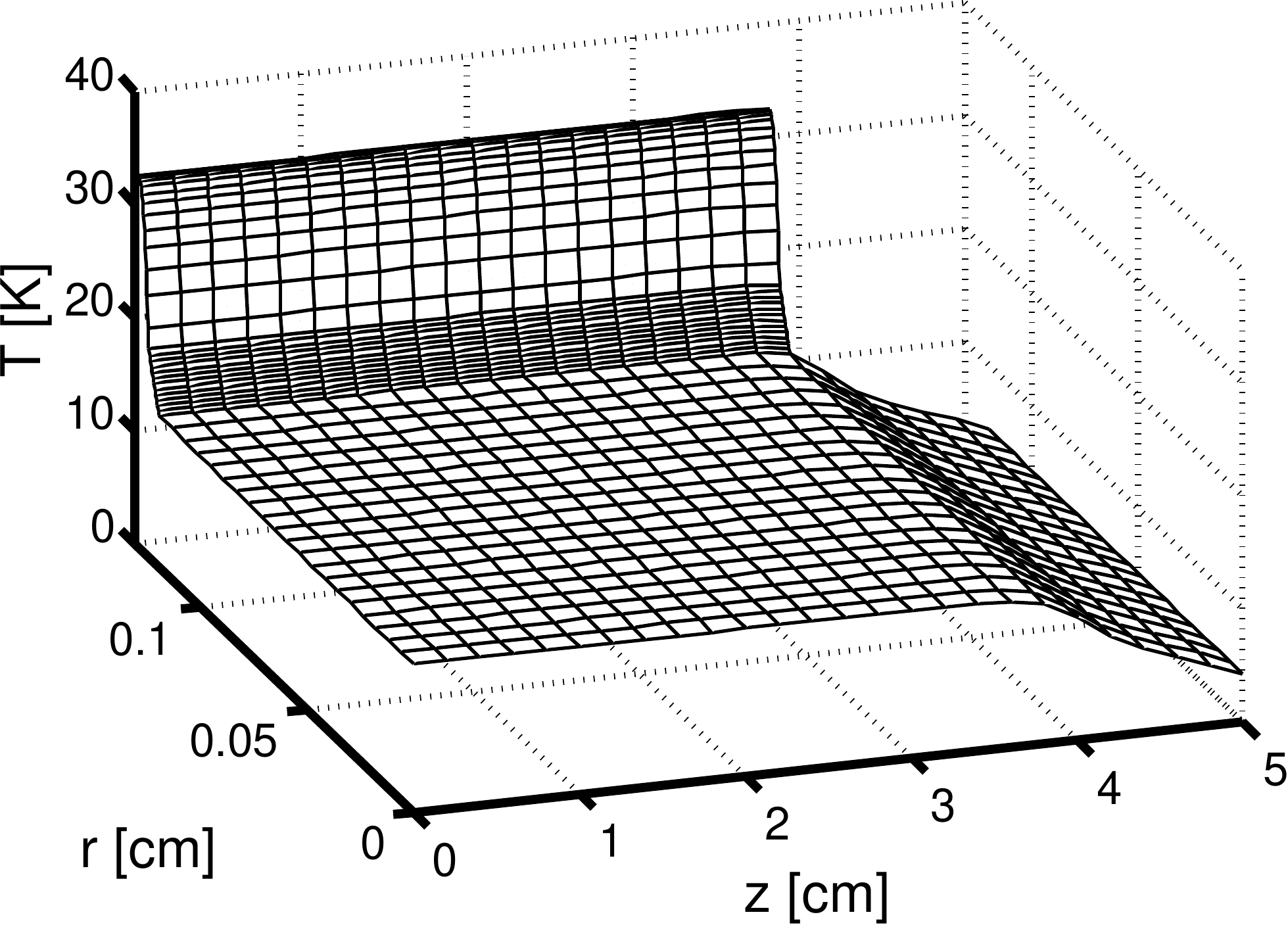}
    \caption{\footnotesize At $t=1\unit{ms}$ after switching on the heat\\ source at the very beginning.}
    \label{fig_temperature3d:heated}
  \end{subfigure}%
  \begin{subfigure}[b]{0.45\textwidth}
    \includegraphics[width=\textwidth]{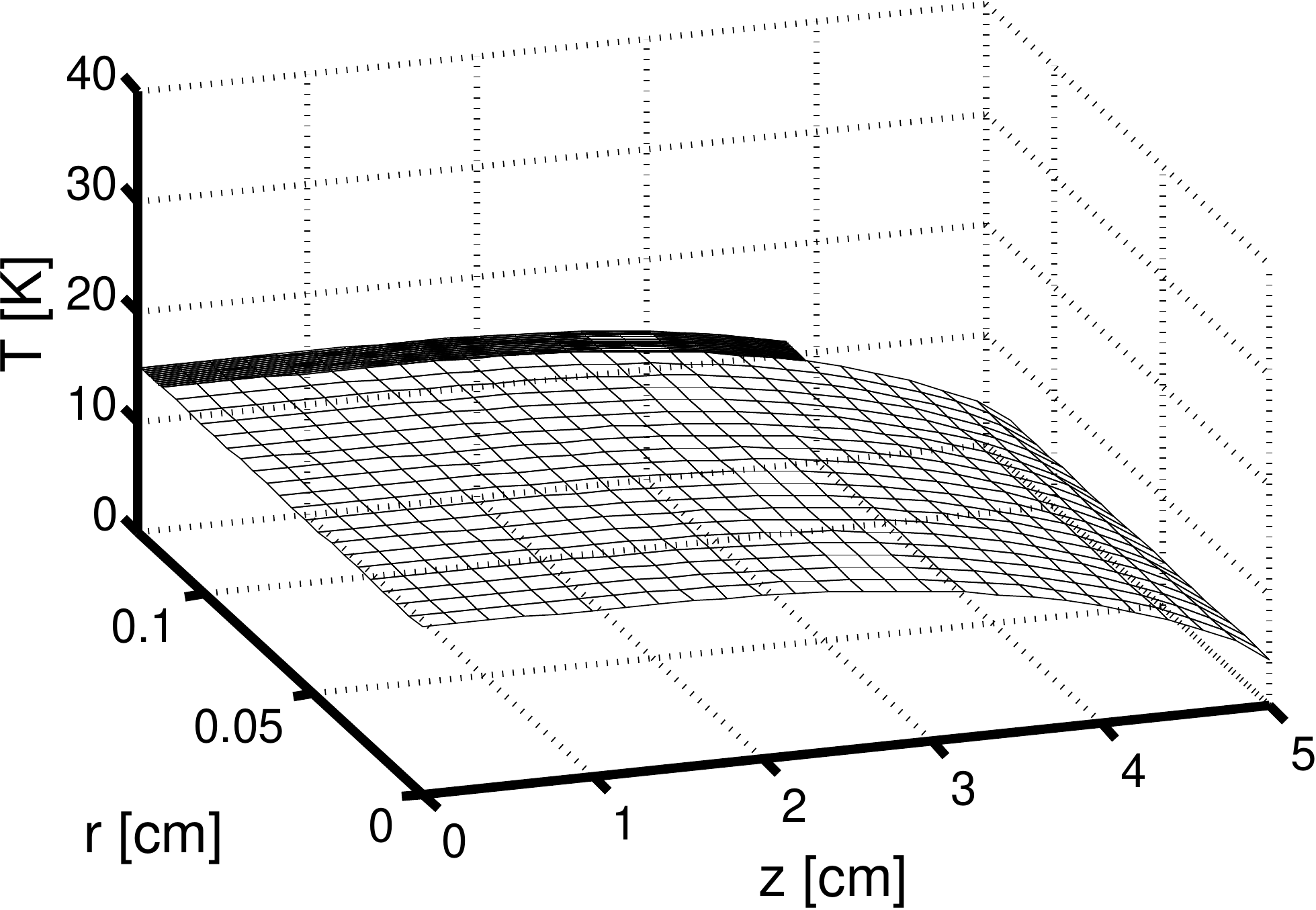}
    \caption{\footnotesize At $t=25\unit{ms}$ before switching on the heat\\ source of the next period.}
    \label{fig_temperature3d:relaxed}
  \end{subfigure}
  \caption{\footnotesize Temperature profile $T(r,z)$ at different times.}
  \label{fig_temperature3d}
\end{figure}

In Fig.~\ref{fig_temperature3d} we show the temperature profiles at the very beginning. The temperature inside the object at $t=1\unit{ms}$ is shown in the left (Fig.~\ref{fig_temperature3d:heated}). At the same moment, the source is switching off in the first period of source function. Because the radius of the cylinder is much smaller than its length, the heat first flows to the central axis and then to the right border, where the cryostat  (the liquid helium temperature terminal) is located at $z_{\mathrm{max}}=5\unit{cm}$.

\begin{figure}[ht!]
  \centering
  \begin{subfigure}[b]{0.45\textwidth}
    \includegraphics[width=\textwidth]{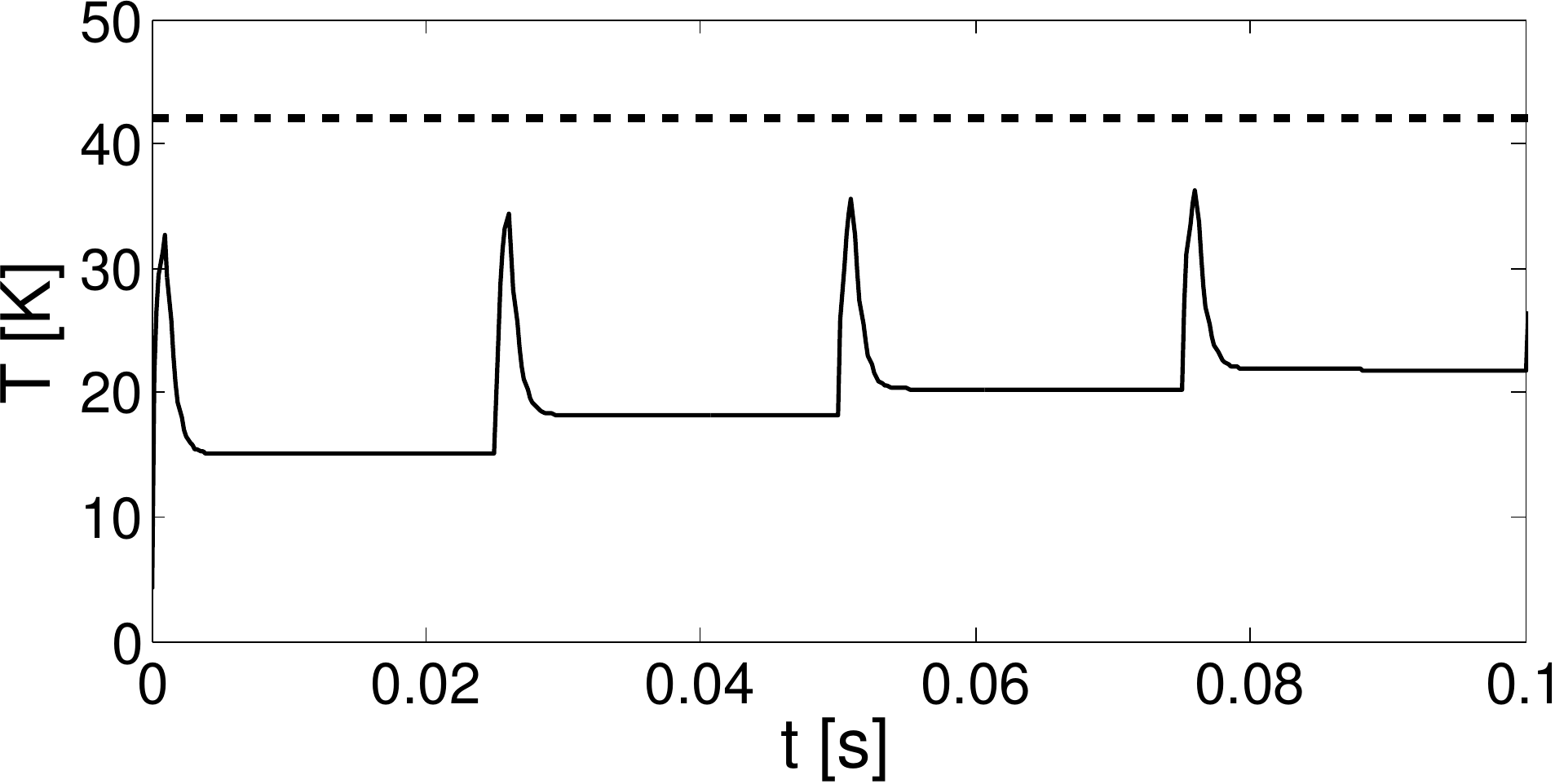}
    \caption{\footnotesize \color{black} The temperature evolution for first four heating pulses.}
    \label{temp_evolution:a}
  \end{subfigure}%
  \hspace{0.5cm}
  \begin{subfigure}[b]{0.45\textwidth}
    \includegraphics[width=\textwidth]{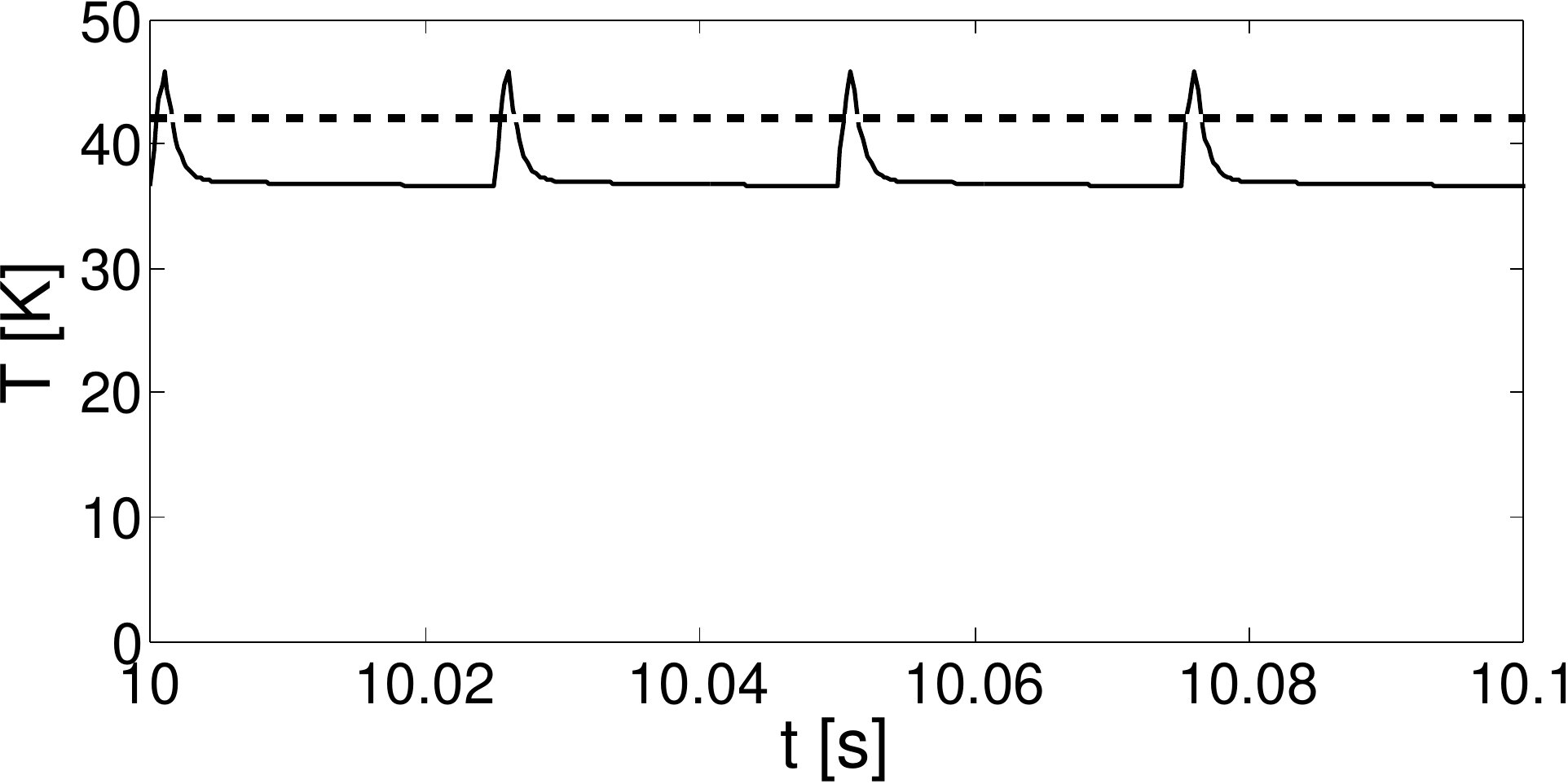}
    \caption{\footnotesize \color{black} The temperature evolution for four heating pulses when requared temperature regime is~achieved.}
    \label{temp_evolution:b}
  \end{subfigure}
\\
  \centering
  \begin{subfigure}[b]{0.45\textwidth}
    \includegraphics[width=\textwidth]{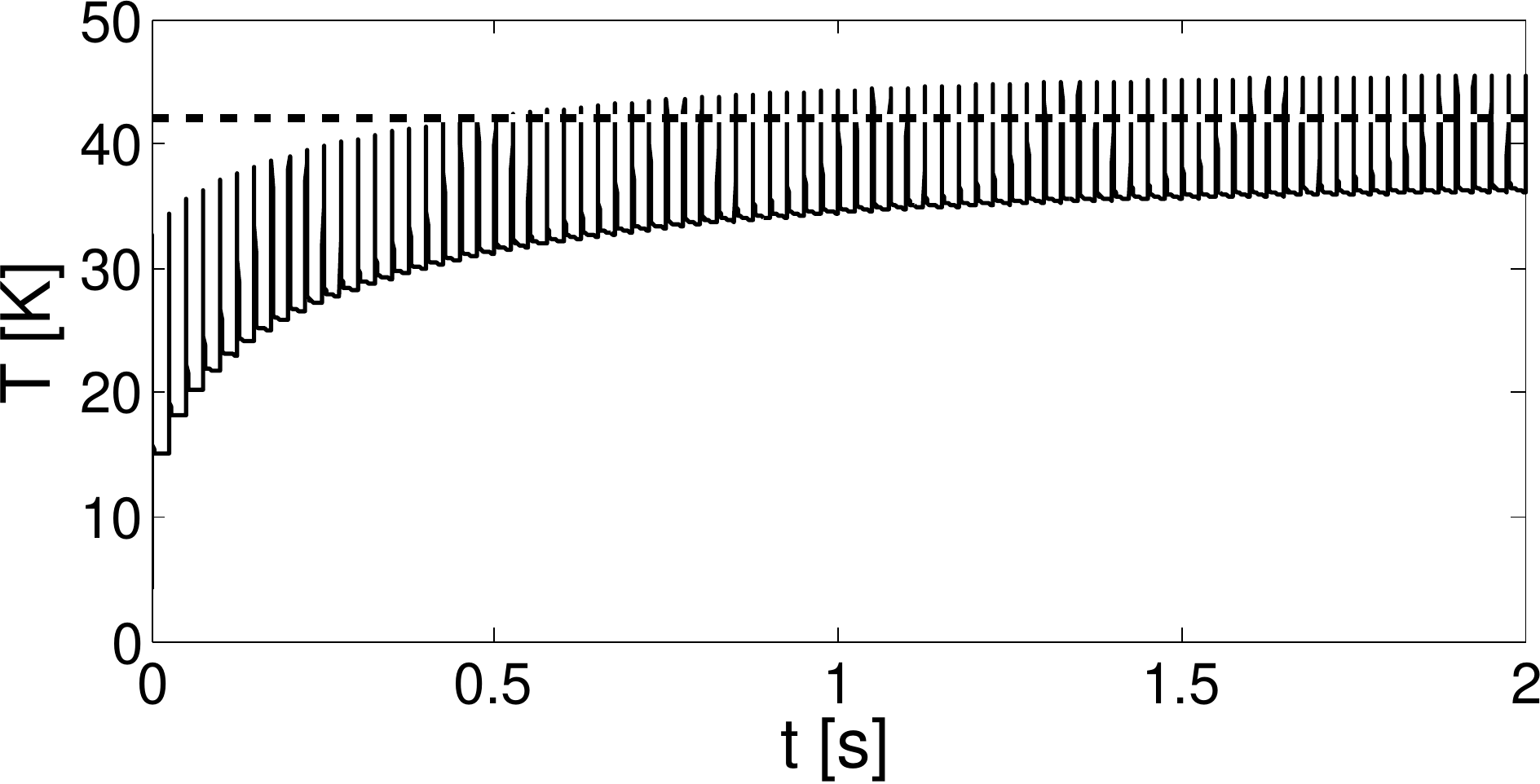}
    \caption{\footnotesize \color{black} The temperature evolution at the beginning up~to~$t=2~\unit{s}$.}
    \label{temp_evolution:c}
  \end{subfigure}%
  \hspace{0.5cm}
  \begin{subfigure}[b]{0.45\textwidth}
    \includegraphics[width=\textwidth]{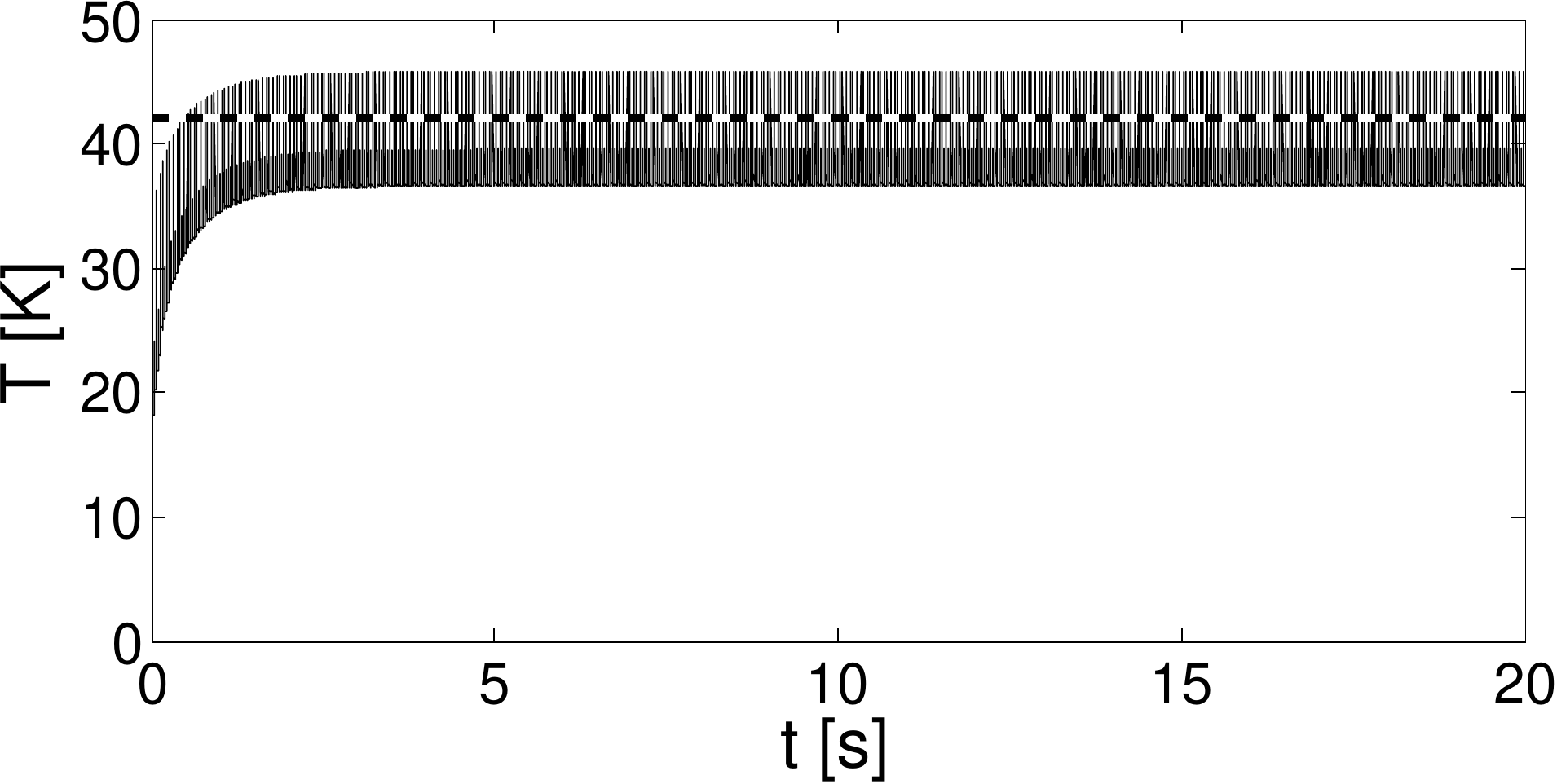}
    \caption{\footnotesize \color{black} The temperature evolution for all period of simulation.}
    \label{temp_evolution:d}
  \end{subfigure}
\caption{\footnotesize \color{black}{The} evolution of the surface temperature at $\left(r=r_{\mathrm{max}}, z = 0\right)$ (solid line). The critical temperature for evaporator and condensation of working gas (dashed line).}
\label{temp_evolution}
\end{figure}

\begin{figure}[ht!]
  \centering
  \includegraphics[width=0.4\textwidth]{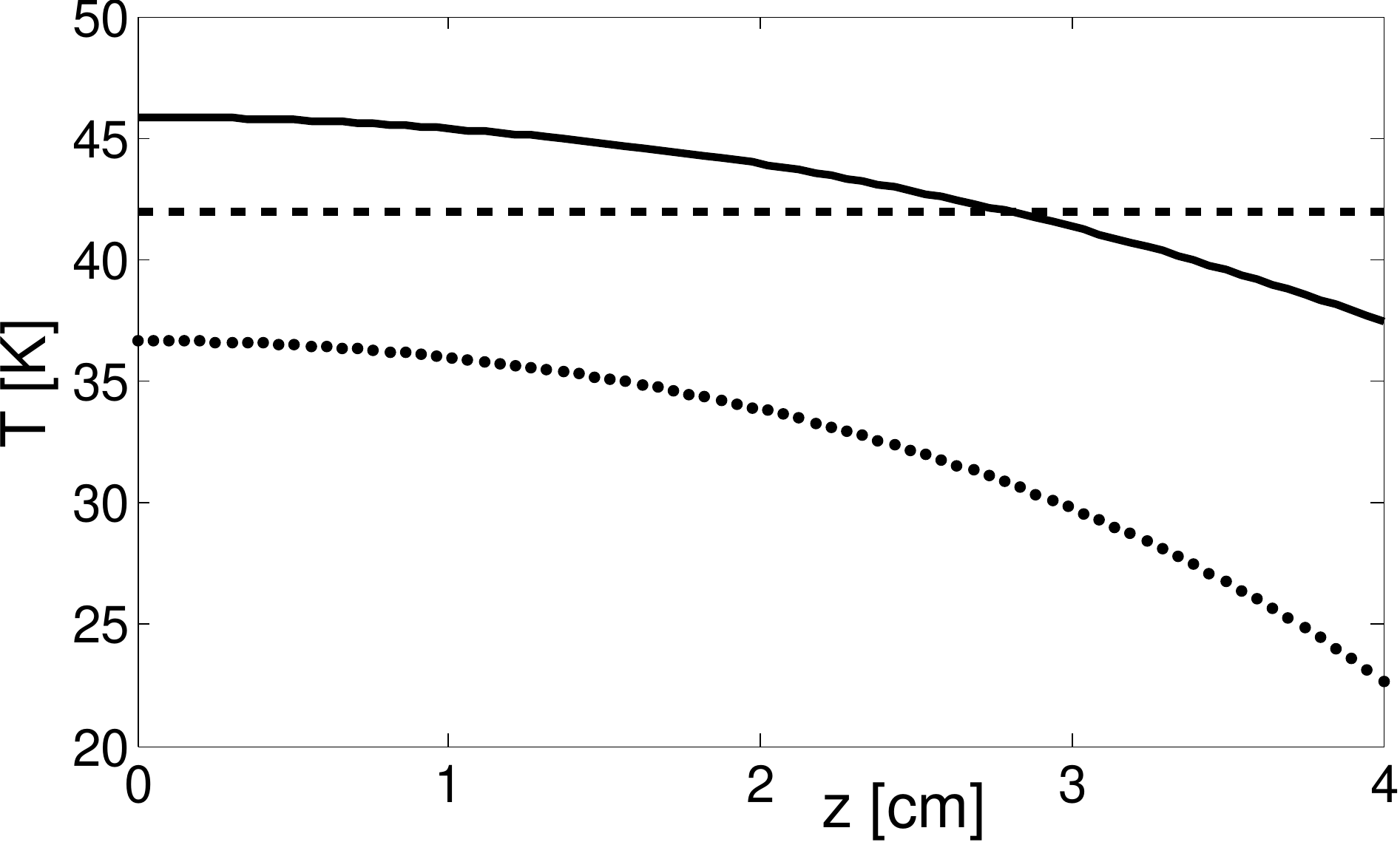}
  \caption{\footnotesize {\color{black} The surface temperature profiles ($r=r_{\mathrm{max}}$) at the stable regime just before at moments of source switching on (dot line) and switching off (solid line). Dashed line is the critical temperature.}}
  \label{fig_temperature}
\end{figure}

In the Fig.~\ref{fig_temperature3d:relaxed} we show the temperature distribution at $t=25\unit{ms}$ (after first period). One period of the source switching $T_{\mathrm{prd}}$ is the sufficient time to equilibrate the temperature in radial direction (see Fig.~\ref{fig_temperature3d:relaxed}). On the other hand, the period of time is not sufficient to relax the temperature to the value right before switching on the source (see Figs.~(\ref{temp_evolution:a})--(\ref{temp_evolution:c})). Such behavior repeats for approximately $10\unit{sec}$ (setting time) until a stable periodic regime is achieved (see Fig.~\ref{temp_evolution:d}). The~setting time varies depending on the design of the object and switching time of the source.

For applications it is very important that the maximum value of the temperature $T_{\mathrm{max}}$ achieves higher value than the critical value of evaporation of working gases ($T_{\mathrm{crit}}\simeq 42.2\unit{K}$) and is distributed along almost the whole working surface ($r=r_{\mathrm{max}}$, $z\in\left[0, z_{0}\right]$) (Fig.~\ref{fig_temperature}). The maximum value of the minimal temperature in the stable regime is less than the critical one (Fig.~\ref{fig_temperature}), as it is required.

Moreover, the width of the temperature peaks is approximately $1$--$1.5\unit{ms}$ and is much smaller in comparison with the whole period (Fig.~\ref{temp_evolution:b}), which allows for working gases to condensate on the surface before the next evaporation.

In the Fig.~\ref{fig_temperature} it is easy to see, that in our particular case, approximately $75\%$ of the working surface is actually useful. This percentage can be controlled, for example, with a proper choice of the amplitude of electric current $I_{0}$. In principle, this equation can be formulated mathematically as source control problem, which can help to raise the efficiency.

\subsection{Results of the OpenCL Implementation}
\label{parall}

The time calculations for different $N \times M$ are given in the Table~\ref{calc_table} (here and in the table: $N = \max\limits_{\forall j}N_j$ and $M = \max\limits_{\forall i}M_i$). To demonstrate the results of OpenCL algorithm, the calculations have been carried out for temperature evolution from $t=0\unit{s}$ up to $t=0.025\unit{s}$ with time step $\tau=2.5 \times 10^{-7}\unit{s}$. In the table we use the following notations: CPU -- Intel Xeon E5-2695 and GPU -- nVidia Tesla K40s. During the compilation of programs -O3 optimization flag has been used.
The calculations have been done on the cluster HybriLIT~\cite{hybrilit}. 
In the Table~\ref{calc_table} we compare the calculation times for CPU and GPU and the speedup using $6$ different grids with the same $N=631$ and various $M$.
It~is~shown that for the very rare grid the calculation time for both CPU and GPU are the same, however, there is an interval of increasing number of points of discretization in axial direction ($M$), where the calculation time using GPU remains the same. Therefore, the choice of our algorithm allows us to increase the density of our computational grid in the axial direction, practically without loss of calculation time.
\begin{table}[ht]
\begin{center}
\caption{Calculation time of OpenCL implementation for different grid size.}
\label{calc_table}
\begin{tabular}{ l c c c c c c}
ine
$N \times M$  &   $631 \times 201$  &   $631 \times 401$  &   $631 \times 601$  &   $631 \times 801$  &   $631 \times 1001$  &   $631 \times 1201$    \\
ine
$T_{\unit{CPU}}$, min.  &  13.0  &  28.8  &  40.5  &  44.9  &  54.9  &  64.0    \\
$T_{\unit{GPU}}$, min.  &  12.9  &  14.1  &  13.0  &  13.0  &  13.0  &  12.9    \\
$T_{\unit{CPU}} / T_{\unit{GPU}}$  &  1.009  &  2.043  &  3.107  &  3.454  &  4.226  &  4.951    \\
ine
\end{tabular}
\end{center}
\vspace{-3mm}
\end{table}

{\color{black}
\section{Summary and Conclusions}
\label{conc}

We have suggested a model of temperature evolution for a multilayer cylindrical object, {\color{black} the} cryogenic cell for pulse injection (in millisecond range) of the gaseous working species into the working space of the ion source.

The algorithm for simulation of heat conduction process with a periodical source in cylindrical multilayer object and its OpenCL realization have been developed. It is based on explicit--implicit method. {\color{black} The} choice of the difference scheme is suitable for the technical realization {\color{black} (because of the structural features of the cryogenic cell)}. For the stability of simulations, the transient process model for source switching is introduced.

It was shown that the temperature regime in the cryogenic cell can be conditionally divided in two parts. The first part is the setting mode (around $10\unit{s}$), {\color{black} in which the required temperature regime} at the surface of the cell has not been reached yet. The second part is the working mode, {\color{black} in which the required temperature regime} at the surface has already been reached.

The mean value of the temperature over the surface changes around the critical one periodically in the interval of $32\unit{K} \lesssim T \lesssim 43.5\unit{K}$ ($T_{\mathrm{crit}}=42.2\unit{K}$). The key characteristics of the cell has been achieved by a particular choice of the model parameters, i.e.~the conducting (heating) layer material, its {\color{black} thicknesses} and source characteristics.

We can conclude that:
\begin{itemize}
\item[---] The introduced multilayer object (the cryogenic cell) could be {\color{black} a} real alternative to mechanical gate valves for the millisecond pulse injection.
\item[---] The suggested algorithm allows to achieve stable solution for all process even {\color{black} in} case of {\color{black} the} fast oscillation of source.
\item[---] {\color{black} The} time of setup mode is much {\color{black} longer} than {\color{black} a} period of {\color{black} the} injection time. {\color{black} This} feature has to be taken into account in future design and {\color{black} operation} of the cryogenic cell.
\item[---] The simulation {\color{black} shows that} the optimal choice of the cell layer materials, thicknesses and source characteristics {\color{black} provides} the required pulse temperature regime {\color{black} on} $75\%$ of the working surface of the cryogenic cell.

\end{itemize}
As an outlook one can mention that further optimization could be done with a variation of the materials and geometry of the cell.
}

\section*{Acknowledgements}
Authors thank Dr.~Edik Ayryan (JINR), Dr.~J\'{a}n Bu\v{s}a, Dr.~Imrich Pokorn\'{y} (TU of Ko\v{s}ice, Slovakia), Mark A. Kaltenborn, and Prof.~David Blaschke (University of Wroc{\l}aw, Poland) for useful advice and technical help. The research was supported by JINR grant No.~14-602-01 and RFBR grants No.~14-01-00628 and No.~14-01-31227.

\bibliographystyle{elsarticle-num}



\end{document}